\documentstyle[aps,pre,graphicx,amssymb,multicol,amsmath,bm]{revtex}%

\begin{document}

\title{Tilt order parameters, polarity and inversion phenomena in
smectic liquid crystals.}

\author{P. K. Karahaliou$^1$, A. G. Vanakaras$^2$
and D. J. Photinos$^1$}

\address{
$^1$ Department of Physics, University of Patras, Patras
26500, Greece. \\
$^2$ Department of Materials Science, University of
Patras, Patras 26500, Greece. }




\maketitle

\begin{abstract}
The order parameters for the phenomenological description of the
smectic-{\it A} to smectic-{\it C} phase transition are formulated on
the basis of molecular symmetry and structure. It is shown that,
unless the long molecular axis is an axis of two-fold or higher
rotational symmetry, the ordering of the molecules in the
smectic-{\it C}  phase gives rise to more than one tilt order
parameter and to one or more polar order parameters. The latter
describe the indigenous polarity of the smectic-{\it C} phase,
which is not related to molecular chirality but underlies the
appearance of spontaneous polarisation in chiral smectics. A
phenomenological theory of the phase transition is formulated by
means of a Landau expansion in two tilt order parameters (primary
and secondary) and an indigenous polarity order parameter. The
coupling among these order parameters determines the possibility
of sign inversions in the temperature dependence of the
spontaneous polarisation and of the helical pitch observed
experimentally for some chiral smectic-{\it $C^{\ast}$} materials.
The molecular interpretation of the inversion phenomena is
examined in the light of the new formulation.
\end{abstract}


\begin{multicols}{2}

\section{Introduction}
The molecular physics of smectic liquid crystals has been
receiving much attention, mainly in connection with
ferroelectricity-related applications of some of these
materials~\cite{bib1}. Nearly three decades ago it was predicted
by R.B. Meyer, and soon afterwards proven experimentally, that a
tilted smectic-$C$ phase made of chiral molecules can exhibit
spontaneous electric polarisation $P_{S}$ within each smectic
layer~\cite{bib2}. This chiral phase, the Sm-$C^{\ast}$, further
differs from the achiral (Sm-$C$, handedness symmetric) phase in
that the azimuthal angle of the director varies linearly with the
distance along the layer normal, thus defining a helical
configuration of definite handedness and constant pitch. Normally,
$P_{S}$ disappears on heating to the non-tilted (orthogonal)
smectic-$A$ (Sm-$A$) phase. More recently, spontaneous
polarisation was detected in a special class of achiral compounds
with bent structure (banana-shaped molecules) forming smectic
phases with form chirality~\cite{bib3,bib4,bib5}.

In general, $P_{S}$ depends very sensitively on the structure of
the molecules that form the Sm-$C^{\ast}$ phase. There are
numerous examples of dramatic changes in $P_{S}$ caused by only
slight modifications of the molecular structure. Usually, the
magnitude of $P_{S}$ increases on lowering the temperature from
the Sm-$A$--Sm-$C^{\ast}$ transition point $T_{A-C^{\ast}}$. A
number of Sm-$C^{\ast}$ compounds, however, deviate from this
behaviour in that the magnitude of $P_{S}$ increases up to some
value from which it decreases on further reducing the temperature.
In several known cases the decrease of $\vert P_{S}\vert$ with
temperature continues until an ``inversion temperature'' $T_{inv}$
is reached at which $P_{S}$ vanishes completely; below that
temperature $P_{S}$ grows again monotonously but with the opposite
sign~\cite{bib6,bib7,bib8,bib9,bib10,bib11}. The temperature
dependence of the tilt angle and of the helical pitch does not
show any particular irregularity around the inversion temperature
of $P_{S}$. Certain compounds exhibit inversion of the pitch
handedness~\cite{bib12} with temperature but this does not seem to
be directly correlated with the sign inversion of $P_{S}$. In one
known case where both types of inversion are exhibited by the same
compound, the two inversions happen at different
temperatures~\cite{bib13}. A sign inverting behaviour of $P_{S}$
with temperature has also been observed in side-chain
Sm-$C^{\ast}$ polymers~\cite{bib14}. A similar sign inversion is
observed, as a function of concentration, in mixtures of achiral
smectic molecules with chiral dopants~\cite{bib15}.

One interpretation proposed for the sign inversion of $P_{S}$ with
temperature is based on the mechanism of competing conformations
that produce opposite contributions to the spontaneous
polarisation~\cite{bib7,bib13}. Another interpretation assumes
competing effects originating from the polar and the quadrupolar
biasing of rotations around the long molecular axis and attributes
the sign reversal to a special case of the coupling of the tilt to
the rotational biasing~\cite{bib16}. Although these two
interpretations address different molecular features, namely
conformational changes and transverse interactions, they are not
necessarily mutually exclusive. In fact, the possibility of sign
reversal of $P_{S}$ is directly obtained in a molecular theory of
primitive smectic molecules consisting of a rigid mesogenic core
and two pendant chains that can rotate about the core axis, thus
producing different conformations~\cite{bib17,bib18}. This theory
explicitly shows that the tilt angle of the core segments is in
general different from the tilt of the chains and that the sign
inversion of $P_{S}$ is related to the variation of this
difference. The variation is driven by packing correlations
between the molecular orientations and the conformations. Such
correlations affect both the conformational sampling and the
sampling of transverse intermolecular interactions that produce
the rotational bias of a given conformer around the long molecular
axis.

Notably, each of the above interpretations implies molecular
features, such as biaxiality and flexibility, that are clearly
beyond the uniaxial rod idealisations underlying the simplest
microscopic and phenomenological descriptions of tilted smectics.
The consideration of more realistic molecular structures is
necessary not only for the interpretation of special phenomena,
such as the inversion of the spontaneous polarisation or of the
pitch, but also for the understanding of more common and
fundamental aspects of smectics, such as micro-segregation. The
latter originates from the chemical differentiation of the two
basic components of the smectic molecules, namely the relatively
rigid mesogenic core and the aliphatic end-chains.
Micro-segregation is the mechanism that drives the formation of
the smectic layers and, combined with the non-linear (zigzag,
bent, etc) structure of the molecules, gives rise to tilt and
polar ordering~\cite{bib17,bib18}. In common smectic molecules,
molecular flexibility consists mainly of internal flexibility of
the end chains and of the possibility of rotations of the chains
as a whole relative to the core. In the absence of any
site-specific interactions, flexibility alone could produce
microsegregation as it is entropically favourable for the
``fluid'' chains to group together~\cite{bib19}. Furthermore, as a
result of the internal relative motions of the submolecular
segments, the average disposition of the flexible, asymmetric,
molecule in the tilted smectic phase cannot in general be
described by a single tilt angle (equivalently, by a single
``director''). Different segments of the molecule could exhibit
different tilt angles with respect to the layer normal.

The existence of more than one director, and associated tilt
angle, has been invoked for the interpretation of the results of
several experimental studies of the Sm-$C$ phase. Deuterium NMR
measurements~\cite{bib20} indicate that different segments of the
flexible smectic molecules do not in general share a common
principal axis (director) of their second rank ordering tensors. A
clear difference between the tilt angles associated with the
mesogenic core and the flexible end--chains of the molecules is
obtained from X-ray measurements~\cite{bib21} in the Sm-$C$ phase.
Analogous conclusions are reached by with IR spectroscopy
by~\cite{bib22}. It is also well known that X-ray measurements
give in general different values for the tilt angle than optical
measurements, indicating that the tilt determined from molecular
packing within the layers need not coincide with the deviation of
the principal optical axis from the layer normal. Such
considerations are consistent with recent results from combined
X-ray and optical studies on ferroelectric liquid crystal
cells~\cite{bib23}. In fact, a single tilt angle description is
strictly applicable only to molecules of uniaxial symmetry.

The same implications of molecular asymmetry and flexibility are
carried over to the polar order parameters; they are in general
different for different segments of the molecule. This is directly
demonstrated by atomistic calculations of the segmental order
parameters~\cite{bib24}. It is also in accord with the observed
sensitivity of the spontaneous polarisation of some categories of
Sm-$C^{\ast}$ compounds to mere changes of the position of the
electric dipole moment within the molecular
frame~\cite{bib1,bib24}.

This paper is concerned with the incorporation of molecular
symmetry and flexibility in the phenomenological description of
the Sm-$A$--Sm-$C$ phase transition. The resulting formulation is
used to analyse the sign inversion of the spontaneous polarisation
and of the pitch observed in some Sm-$C^{\ast}$ materials since
such phenomena are thought to reflect particular effects of
molecular structure and conformation on smectic ordering.

The next section deals with the identification of the relevant
order parameters of the Sm-$A$ and Sm-$C$ phases in relation to
molecular structure and symmetry. These order parameters are then
used in section III to formulate a Landau expansion of the free
energy of the Sm-$A$--Sm-$C$ transition. The new Landau expansion
is used to derive the temperature dependence of the order
parameters and therefrom to investigate the conditions leading to
sign inversion of the parameters associated with the spontaneous
polarisation, in section IV, and the handedness of the pitch in
the Sm-$C^{\ast}$ phase, in section V. The new description is
compared with the conventional Landau expansion in section VI.

\section{Symmetries and order parameters.}
The Sm-$C$ phase has a mirror symmetry plane, the ``tilt plane'',
perpendicular to the layers and a twofold rotation symmetry axis
($C_{2}$) in the direction normal to the tilt
plane~\cite{bib1,bib2}. The intersection of the twofold axis with
the mirror plane defines the centre of inversion symmetry of the
phase. In what follows, the layer normal is identified as the
$Z$-axis of a phase-fixed reference frame and the twofold symmetry
axis is taken to be the $X$-axis of the frame. Accordingly, the
above symmetries imply invariance of the molecular probability
distribution, and thereby of the free energy of the phase, with
respect to the following two transformations,
\begin{equation}
\label{eq1}
 X \to  - X\text{ (plane of symmetry)} \quad {\rm ,}
\end{equation}
\begin{equation}
\label{eq2}
 (Y,Z) \to ( -Y, -Z)\text{ (twofold rotation)} \quad{\rm .}
\end{equation}

As a result of the invariance with respect to these
transformations the phase is also invariant with respect to change
of handedness of the $XYZ$ frame, i.e. achiral. The
symmetry of the Sm-$A$ phase differs in that it is invariant
separately with respect to
\begin{equation}
\label{eq3}
Y \to  - Y\text{ and }Z \to  - Z \quad {\rm .}
\end{equation}

\subsection{Uniaxial molecules}
\begin{figure}[t]
  \includegraphics*[width=8.6cm]{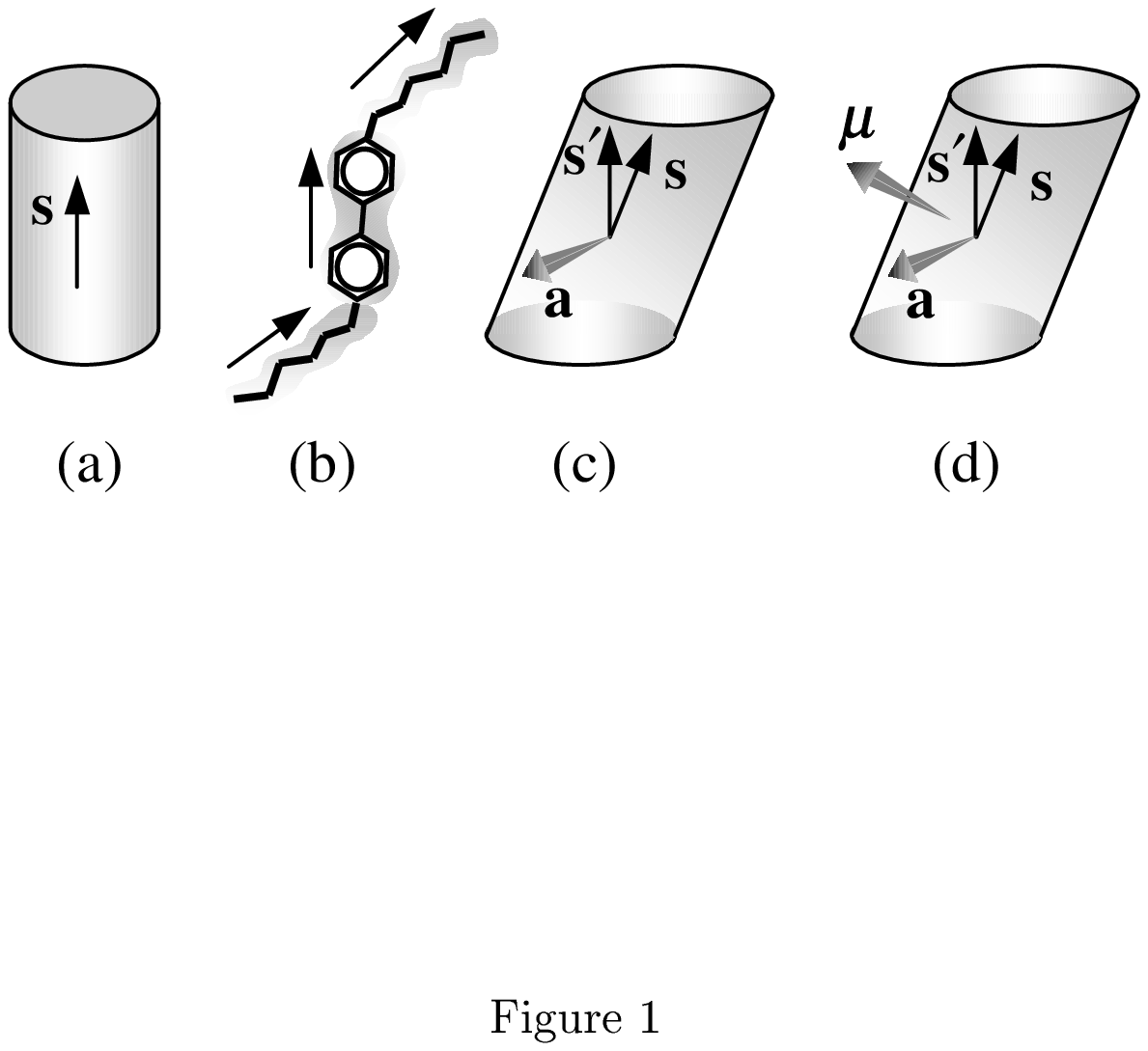}\\
\caption {(a) Right cylinder representing the molecular symmetry
of rigid uniaxial molecules. The unit vector \textbf{s} defines
the direction of the axis of full rotational symmetry of the
molecule. (b) Coarse representation of the generic structure of
real smectic molecules in the most symmetric case. The molecules
have a plane of (statistical) symmetry (mirror plane, coinciding
with the plane of the drawing), an axis of two-fold rotational
symmetry (perpendicular to the symmetry plane) and an inversion
center (at the point where the two-fold axis intersects the mirror
plane). The three arrows represent the vectors describing the
direction of the mesogenic core and of the axes of the two end
chains in their most extended conformation. (c) Oblique cylinder
representation of a molecule bearing the same symmetries as in (b)
but disregarding other structural and conformational features. The
two unit vectors $\mathbf{s}$ and $\mathbf{s'}$ are rigidly
attached to molecule; they specify its orientation and define its
mirror symmetry plane. The pseudovector $\mathbf{a} =
\mathbf{s}\times \mathbf{s'}$ is normal to the symmetry plane. (d)
A dipole moment $\bm{\mu}$ is attached to the oblique cylinder in
(c). If $\bm{\mu}$ has a non-vanishing component in the direction
of $\mathbf{a}$, the attachment of the dipole leads to the
breaking of the mirror symmetry. The molecule would then become
chiral with respect to its electrostatic interactions.}
 \label{fig1}
\end{figure}

If the molecules forming the smectic phase are approximated by
uniaxially symmetric rigid objects, then the orientation of each
molecule is specified in terms of a single unit vector
$\mathbf{s}$ along the molecular axis of full rotational symmetry
(see Fig.~\ref{fig1}(a)). In that case, the orientational order
parameters, i.e. the thermal averages of tensors of various ranks
that can be formed form the components of $\mathbf{s}$, reflect
the symmetries of the Sm-$C$ phase in the following way: \\
The zero rank (scalar) order parameters are trivial
$({\left\langle{\mathbf{s}^{2}}\right\rangle}=1)$ in view
of the assumed rigidity of the molecules.\\
The first rank (vector) order parameters vanish in
view of Eqs. (\ref{eq1}) and (\ref{eq2}),
\begin{equation}
\label{eq4}
\left\langle {s_X } \right\rangle ,\left\langle {s_Y }
\right\rangle ,\left\langle {s_Z } \right\rangle  = 0 \quad {\rm ,}
\end{equation}
The second rank order parameters are the components of the
symmetric and traceless tensor
\begin{equation}
\label{eq5}
\eta _{ab}  = (3 < s_a s_b  >  - \delta _{ab} )/2
\quad {\rm ,}
\end{equation}
with the subscripts $a$, $b$ denoting components along the $X$,
$Y$, $Z$ axes. The order parameters $\eta_{XY}$ and $\eta _{XZ}$
vanish as a result of the symmetry transformation in Eq.
(\ref{eq1}). The diagonal components $\eta_{XX}$, $\eta_{YY}$,
$\eta _{ZZ}$ survive both symmetry operations in Eqs. (\ref{eq1}),
(\ref{eq2}) and so does the off diagonal component $\eta_{YZ}$.
The latter gives a measure of the breaking of the rotational
symmetry about the layer normal ($Z$-axis) due to the tilted
ordering of the molecules. The additional symmetry, Eq.
(\ref{eq3}), of the Sm-$A$ phase leads to $\eta_{YZ}=0$. Thus
the primary order parameter for the distinction between the Sm-$A$
and the Sm-$C$ phase is $\eta_{YZ}$.

This order parameter is often expressed in terms of the ``tilt
pseudovector''
\begin{equation}
\label{eq6}
 \mathbf{t} = (\mathbf{Z} \times \mathbf{\tilde
Z})(\mathbf{Z} \cdot \mathbf{\tilde Z}) \quad {\rm ,}
\end{equation}
where $\mathbf{\tilde Z}$ is the unit vector in the direction of
the principal axis of the ordering tensor $\eta _{ab} $. The
principal axis frame $\tilde{X} \tilde{Y} \tilde{Z}$ is
obtained by rotating the $XYZ$ frame about the $X$-axis by an
angle $\theta$ such as to diagonalise the tensor $\eta _{ab}$,
i.e. to obtain the frame for which $\eta_{\tilde Y\tilde Z}= 0$
or, equivalently, for which the order parameter $\eta_{ZZ}$
acquires its maximum value $\eta_{\tilde{Z}\tilde{Z}}$. Obviously
$\tilde{X}$ coincides with $X$ and the pseudovector $\mathbf{t}$
is along $X$. The angle $\theta$ is related to the tilt
pseudovector and to the components of $\eta_{ab}$ as follows
\begin{equation}
\label{eq7}
 \sin 2\theta = 2{\rm {\bf X}} \cdot \mathbf{t} =
2\eta_{YZ} / (\eta_{\tilde {Z}\tilde {Z}} - \eta _{\tilde
{Y}\tilde {Y}} ) \quad {\rm .}
\end{equation}
Since the onset of the (achiral) Sm-$C$ phase is marked simply by
the appearance of a non-vanishing value of the tilt vector, the
Landau expansion~\cite{bib1} of the equilibrium free energy
difference describing the Sm-$A$--Sm-$C$ phase transition is an
expansion in the single order parameter $\mathbf{t}$. Furthermore,
due to the thermodynamic equivalence of the states with
$\mathbf{t}$ and $-\mathbf{t}$, the expansion contains only even
powers of $\mathbf{t}$,
\begin{equation}
\label{eq8}
g_{A - C} = {\frac{{1}}{{2}}}at^{2} + {\frac{{1}}{{4}}}bt^{4} +
\ldots \quad {\rm .}
\end{equation}
This is the conventional form of the Landau expansion for the
Sm-$A$--Sm-$C$ transition in the absence of external
fields~\cite{bib25,bib26,bib27,bib28}. It clearly does not involve
any kind of polarity order parameter. It should be recalled,
however, that this description is valid only under the assumption
that a single vector $\mathbf{s}$ is sufficient to describe the
molecular orientation, or equivalently, that the molecules in the
smectic phase behave as rigid uniaxial objects.

\subsection{Minimal deviation from uniaxial molecules}
Real smectic molecules are of course flexible, their shape is not
uniaxial and their orientation within the smectic phase cannot be
fully specified by a single vector (see Fig.~\ref{fig1}(b)). In
fact the complete specification of the orientation and
conformation of the molecule requires at least as many unit
vectors as there are molecular segments capable of moving relative
to one another. In what follows we demonstrate that the
description of the Sm-$A$--Sm-$C$ phase transition becomes
qualitatively different if one goes beyond the uniaxial
idealization of the molecular structure. This will be done by
minimally extending the single vector description to a description
in terms of two molecular unit vectors but the formulation can be
readily generalized to more complex molecular structures.

Consider a molecular structure such as the one shown in
Fig.~\ref{fig1}(b).In the most symmetrical case the structure is
centrosymmetric, the plane of the fully extended conformation of
the molecule is a mirror symmetry plane and the axis perpendicular
to that plane at the inversion centre of the molecule is a twofold
symmetry axis. Ignoring for the moment molecular flexibility and
structural details, these symmetries are conveyed by the oblique
cylinder of Fig.~\ref{fig1}(c). Unlike the right cylinder of
Fig.~\ref{fig1}(a), the orientation of this object cannot be
completely specified by a single unit vector $\mathbf{s}$. A
second unit vector $\mathbf{s'}$ is required. A convenient choice
of unit vectors $\mathbf{s}$, $\mathbf{s'}$ is shown in
Fig.~\ref{fig1}(c). A measure of the deviation from prefect
rotational symmetry about a single ``long'' axis is then provided
by the pseudovector $\mathbf{a} = \mathbf{s}\times \mathbf{s'}$.
This pseudovector is normal to the symmetry plane of the oblique
cylinder; its direction can be used to differentiate between the
two ``faces'' of the molecule, i.e. the two halves of the molecule
separated by the symmetry plane.

The symmetries of Eqs. (\ref{eq1}) and (\ref{eq2}) imply that the
first rank order parameters associated with the two unit vectors
of the oblique cylinder vanish, $\left\langle \mathbf{s}
\right\rangle=0=\left\langle \mathbf{s'} \right\rangle$. By
analogy with Eq. (\ref{eq5}) there are two second--rank order
parameter tensors, one for each of the vectors
$\mathbf{s},\mathbf{s'}$, namely
\begin{equation}
\label{eq9}
\eta _{ab}^{(s)} = (3 < s_{a} s_{b} > - \delta _{ab} )
/ 2 \quad {\rm ,}
\end{equation}
and
\begin{equation}
\label{eq10}
\eta_{ab}^{({s'})} = (3 < {s'}_{a} {s'}_{b} > -
\delta_{ab} ) / 2 \quad {\rm .}
\end{equation}
There is also a third, mixed, second rank order parameter tensor
\begin{equation}
\label{eq11} \eta _{ab}^{(s,{s'})} = (3 < s_{a} {s'}_{b} +
{s'}_{a} s_{b} >/2 - (\mathbf{s}\cdot\mathbf{s'})\delta_{ab} ) / 2
\quad {\rm .}
\end{equation}
Only the $YZ$ off--diagonal components of these tensors survive
the symmetry operations in Eqs. (\ref{eq1}), (\ref{eq2}). Now, the
diagonalisation of each of the tensors $\eta_{ab}^{(s)}$, $\eta
_{ab}^{({s'})}$, $\eta_{ab}^{(s,{s'})}$ requires in general a
different rotation about the $X$-axis. Accordingly, there are
three different tilt angles $\theta^{(s)}$, $\theta^{({s}')}$,
$\theta ^{(s,{s}')}$ defining three different director frames
(frames of principal axes). The three tilt angles, and the
associated tilt vectors $\mathbf{t}^{(s)}$, $\mathbf{t}^{({s}')}$,
$\mathbf{t}^{(s,{s}')}$ are related to the components of the
respective tensors analogously to Eqs. (\ref{eq6}) and
(\ref{eq7}). As shown in the Appendix, the choice of the three
independent tilt order parameters to represent the breaking of the
rotational symmetry about the layer normal is not unique. It is
also shown there that in the case of perfectly rigid molecules,
one of the three parameters can be eliminated by choosing properly
the molecular frame of axes.

\begin{figure}[t]
  \centering
  \includegraphics*[width=8.6cm]{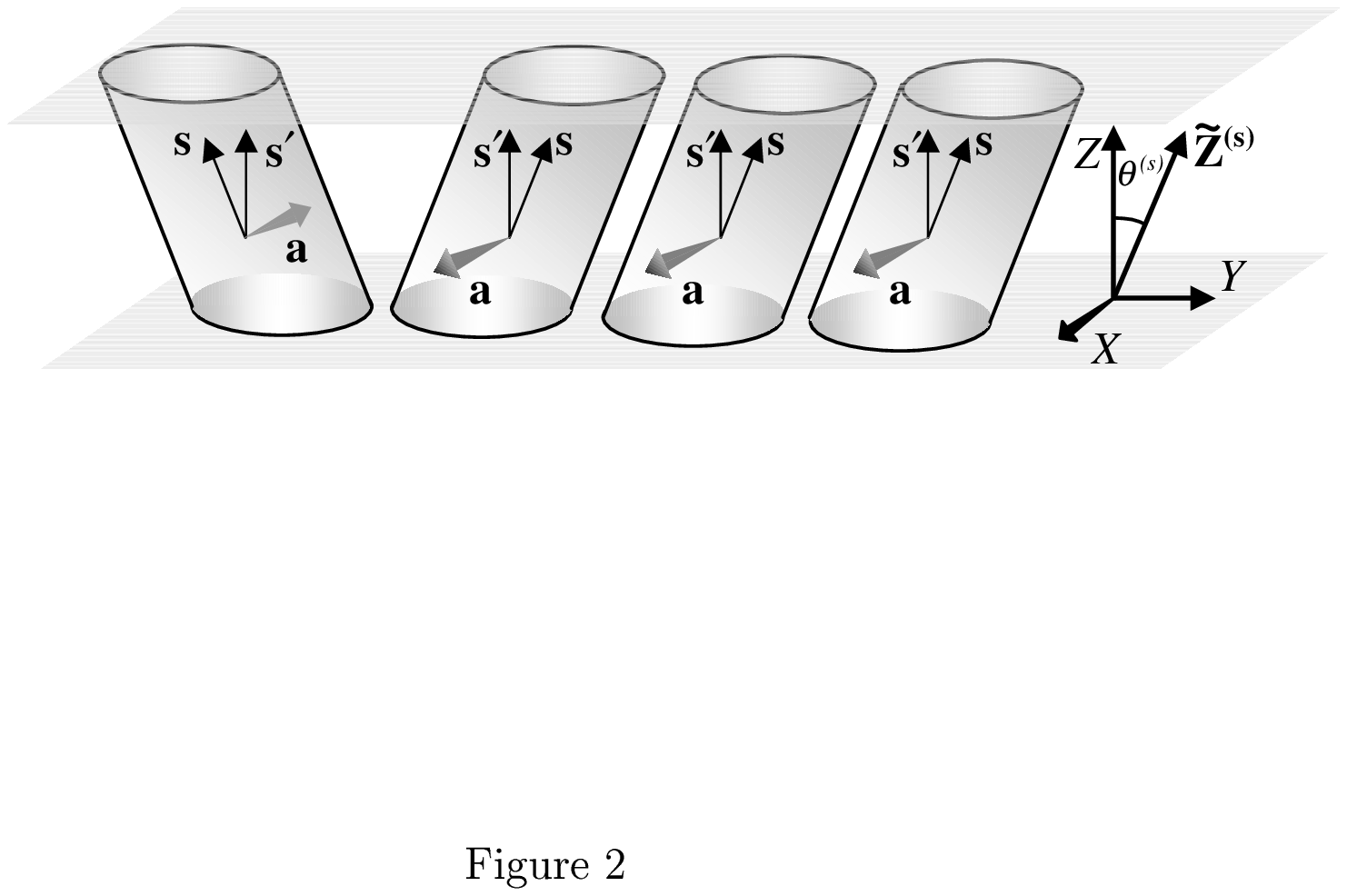}
\caption { Illustration of the packing mechanism giving rise to
polar ordering within a single layer of the tilted phase. With the
director ${\rm {\bf \tilde {Z}}}^{(s)}$ tilted to the right
relative to the layer normal, the statistically dominant molecular
configurations are represented by the three oblique cylinder
molecules on the right. For such configurations, the pseudovector
\textbf{a} points out of the plane of the figure (the ``tilt
plane''). Molecules configured as the oblique cylinder on the left
end have \textbf{a} pointing into the plane of the figure but
these configurations deviate from the preferred tilt direction and
are therefore statistically less favored by the packing
constraints. Accordingly the average projection of the
pseudovector \textbf{a} along the direction normal to the plane of
the figure (the $C_{2}$ axis of the phase) will not vanish. The
value of this projection defines the indigenous polarity order
parameter $\mathbf{P_I}$ of Eq. (\ref{eq13}). The molecules chosen
for this illustration have a plane of mirror symmetry
(perpendicular to \textbf{a}) in order to stress that the polarity
of the tilted phase has nothing to do with molecular chirality.}
\label{fig2}
\end{figure}

The existence of more than one tilt order parameter is not the
only difference from the uniaxially symmetric molecules. Another,
perhaps more important, difference is that the pseudovector
$\mathbf{a}$ singles out a unique transverse molecular direction
and this makes it possible to define the (pseudovector) order
parameter ${\left\langle {\mathbf{a}} \right\rangle}$. The $Y$ and
$Z$ components of ${\left\langle \mathbf{a} \right\rangle}$ vanish
as a result of the symmetry operations of Eqs. (\ref{eq1}) and
(\ref{eq2}), but the $X$ component survives these operations and
therefore the respective order parameter
\begin{equation}
\label{eq12}
 < a_{X} > = < s_{Y} {s'}_{Z} - {s'}_{Y} s_{Z} >
\end{equation}
\noindent acquires a non--vanishing value in the Sm-$C$ phase.
This order parameter describes the indigenous polar
ordering~\cite{bib17} exhibited by the molecules as a result of
the tilted alignment within the smectic layers. The microscopic
origin of the indigenous polarity is depicted in Fig.~\ref{fig2}
for the molecules whose shape can be approximated by the oblique
cylinders of Fig.~\ref{fig1}(c): with the director
$\mathbf{\tilde{Z}}^{(s)}$ tilted to the right of the $Z$-axis and
with the $X$-axis pointing outwards from the plane of the figure,
the combination of stratification and alignment constraints
favours the molecular configurations for which $\mathbf{a}$ points
in the positive direction of the $X$-axis ($\mathbf{a} \cdot
\mathbf{X} > 0$) over those for which \textbf{a} points in the
negative direction ($\mathbf{a} \cdot \mathbf{X} < 0$).
Accordingly, on the average $\mathbf{a} \cdot \mathbf{X}$ will
acquire a positive value $<a_{X}>$.

The existence of the tilt-induced polar ordering was demonstrated
using explicit molecular models of the Sm-$C$ phase taking into
account phase symmetry and orientation-conformation correlations
dictated by the tilted stratified ordering~\cite{bib17,bib18}. It
was also pointed out that this type of polar ordering, the
indigenous polarity, was overlooked in all previous molecular
theories of tilted smectics. For notational convenience, the polar
ordering can be represented by an indigenous polarity
(pseudovector) order parameter $\mathbf{P_I}$ directed along the
$X$-axis of the phase and defined as follows,
\begin{equation}
\label{eq13} \mathbf{P_I}=< \mathbf{a} > /|\mathbf{a}| =
\mathbf{X} < a_X > /|\mathbf{a}| \quad {\rm .}
\end{equation}

\subsection{Spontaneous polarisation and molecular chirality}
Clearly, the indigenous polarity is not a result of molecular
chirality and is present irrespectively of whether the phase
exhibits an electric spontaneous polarisation $\mathbf{P_S}$ or
not. In fact the appearance of a spontaneous polarisation can be
considered as a manifestation of the indigenous polarity when the
smectic molecules possess a permanent dipole moment. For example,
if the oblique cylindrical molecules in Fig.~\ref{fig2} possess a
dipole moment $\bm{ \mu }$, rigidly attached to the molecular
frame (see Fig.~\ref{fig1}(d)) and with a non-vanishing component
along $\mathbf{a}$, then the indigenous polarity would give rise
to a spontaneous polarisation vector $\mathbf{P_S}$ along the
$X$-axis. The spontaneous polarisation $\mathbf{P_S}$ is in this
case expressed in terms of the indigenous polarity order parameter
$\mathbf{P_I}$ according to:
\begin{equation}
\label{eq14}
\mathbf{P_S}  = \left\langle \bm {\mu}  \right\rangle =
\mu_\bot ^ *  \mathbf{P_I}  \quad
{\rm ,}
\end{equation}
\noindent where $\mu_\bot^*  \equiv (\bm{ \mu } \cdot
\mathbf{a})/|\mathbf{a}|$ is the (pseudoscalar) measure of the
``transverse dipole component''. It should be noted at this point
that the presence of a dipole moment with a non vanishing
component along $\mathbf{a}$ breaks the mirror symmetry of the
oblique cylindrical molecule i.e. introduces a chiral asymmetry.
It is clear, however, that this chiral asymmetry does not produce
the polar ordering; it is only involved with the manifestation of
the latter in the form of an electric spontaneous polarisation.

The example of the oblique cylinder was used here as a minimal
deviation from rotationally symmetric molecular structures to
provide a simple concrete illustration of the mechanisms
underlying the relation of polarity to tilted ordering. However,
the basic conclusions reached in this section, namely that the
molecular ordering in the Sm-$C$ phase is intrinsically polar and
not adequately described by just a single tilt order parameter (or
a single ``director''), can be readily carried over to more
realistic examples of molecular structure and flexibility.

\section{Landau expansion.}
Given that there are more than one tilt order parameters and at
least one indigenous polarity order parameter, it is necessary to
describe how these order parameters are incorporated in the
phenomenological Landau expansion of the free energy for the
Sm-$A$--Sm-$C$ phase transition. This is addressed in the present
section using, for simplicity, the example of molecules with the
symmetries of the oblique cylinder.

Since there are three second rank tensors, as in Eqs.,
(\ref{eq9})-(\ref{eq11}), the diagonalisation of which defines
three different tilt vectors, the formulation of a Landau
expansion is not as straightforward as in the case of a single
tilt vector. Obviously, any linear combination of the three
tensors constitutes a new tensor whose diagonalisation defines a
tilt vector. It is thus possible to use in place of the original
three tensors any three linearly independent combinations thereof
(see the Appendix). Of course all such choices are physically
equivalent and the respective expansions can be transformed into
one another. The actual choice is therefore dictated by
considerations of simplicity and physical clarity. As shown in the
Appendix, a description in terms of just two independent tilt
vectors can be obtained in the case of rigid molecules by properly
choosing the molecular frame. The explicit consideration of
several molecular segments and tilt vectors, without restrictions
on flexibility, is treated in detail elsewhere~\cite{bib29}. Here
we consider systems that can be described in terms of two tilt
vectors, $\mathbf{t}$ and $\mathbf{t'}$ both of which acquire non
zero values at the same transition temperature $T_{A-C}$. The tilt
vectors $\mathbf{t}$ and $\mathbf{t'}$ will be referred to as the
primary and secondary tilts respectively. Physically, the primary
tilt could be assigned to represent, for example, the tilted
ordering of the mesogenic core and the secondary tilt to represent
the effective mean tilt of the pendant chains. Another possibility
for the physical content of $\mathbf{t}$ and $\mathbf{t'}$ is to
describe the average tilt of the overall molecule ($\mathbf{t}$)
and the weighted spread in the tilts exhibited by the different
molecular segments ($\mathbf{t'}$). Similarly, a single polarity
parameter $\mathbf{P_I}$ will be used, which is understood to
represent the polar ordering of a unit pseudovector defined by the
vector product of two appropriately chosen molecular vectors.

With all three pseudovectors $\mathbf{t},\mathbf{t'},\mathbf{P_I}$
along the $X(C_{2})$ axis, the rotational invariants that can
enter in the extended Landau expansion are the scalar quantities
$(\mathbf{t} \cdot \mathbf{t})$, $(\mathbf{t'} \cdot
\mathbf{t'})$, $(\mathbf{P_I} \cdot \mathbf{P_I})$, $(\mathbf{t}
\cdot \mathbf{t'})$, $(\mathbf{t} \cdot \mathbf{P_I} )$,
$(\mathbf{t'} \cdot \mathbf{P_I})$. Accordingly, the extended
version of the expansion of Eq. (\ref{eq8}) for the free energy of
a single smectic layer contains the following leading terms
\begin{equation}
\label{eq15}
g_{A - C} = {\frac{{1}}{{2}}}at^{2} + {\frac{{1}}{{2}}}{a'}{t'}^{2}
-
ctP_{I} - {c}'{t}'P_{I} + {c''}t{t'} + {\frac{{1}}{{2}}}dP_{I}^{2} +
{\frac{{1}}{{4}}}bt^{4} + \ldots
\end{equation}
Here the pseudovectors $\mathbf{t},\mathbf{t'},\mathbf{P_I}$ are
replaced, for notational simplicity, by their projections
$t,t',P_I$ along the $X$-axis. The coefficients
$a,a',b,c,c',c'',d$ are all scalar (handedness-symmetric)
quantities. The coefficient $d$ is associated with the decrease in
entropy resulting from the polar ordering within the smectic layer
and is therefore positive ($d > 0$). The signs of the coefficients
$c,{c}',{c}''$, associated with the bilinear coupling
contributions among the parameters $t,{t}',P_{I}$, depend on the
choice of the relative signs of the molecular vectors. The
differentiation between the primary tilt $t$ and the secondary
$t'$ in the expansion is made by the inclusion of a fourth--power
contribution only for the former, with $b > 0$ and by the strong
temperature dependence of the coefficient $a$. The latter
coefficient is assumed to change sign with temperature near the
Sm-$A$--Sm-$C$ transition, whereas the coefficient $a'$ is assumed
to be, as all the other coefficients, slowly varying with
temperature around the transition. For flexible molecules,
however, the variation of the conformational statistics (and
thereby of the ``effective'' molecular structure) with
temperature, could enhance considerably the temperature dependence
of these coefficients. The sign of $a'$ is assumed to be positive,
corresponding to the dominance of the entropy decrease associated
with the secondary tilt over the respective lowering of the
internal energy. Higher order terms have been omitted from the
expansion in Eq. (\ref{eq15}) to avoid excessive mathematical
burden. Terms, however, such as $P_{I}^{2} t^{2}$ could be of
particular importance for the correct description of the
underlying physics and are therefore not negligible in
general~\cite{bib1,bib27}.

Minimisation of the free energy in Eq. (\ref{eq15}) with respect
to $t$, $t'$ and $P_I$ yields the following expressions for
these order parameters in terms of the expansion coefficients
\begin{equation}
\label{eq16}
t^{2} = (h - a) / b \quad {\rm ,}
\end{equation}
\begin{equation}
\label{eq17}
{t}' / t = r \quad {\rm ,}
\end{equation}
\begin{equation}
\label{eq18}
P_{I} / t = R \quad {\rm ,}
\end{equation}
where
\begin{equation}
\label{eq19} h \equiv [c^{2} + (c{c'} - {c''}d)^{2} / ({a'}d -
{c'}^{2})] / d \quad {\rm,}
\end{equation}
\begin{equation}
\label{eq20} r \equiv (c{c}' - {c''}d) / ({a}'d - {c'}^{2}) \quad
{\rm ,}
\end{equation}
and
\begin{equation}
\label{eq21} R \equiv ({a}'c - {c}'{c''}) / ({a'}d - {c'}^{2})
\quad {\rm .}
\end{equation}

It is usually assumed that the strong dependence of $a$ on
temperature near the phase transition is adequately described by
the form
\begin{equation}
\label{eq22}
a / b \approx a_{1} (T - T^{0}) \quad {\rm ,}
\end{equation}
with $a_{1} $ constant and positive and $T^{0}$ a characteristic
temperature constant. For $T$ near $T^{0}$, $h/b$ can be
approximated by
\begin{equation}
\label{eq23}
h / b \approx h_{0} + h_{1} (T - T^{0}) \quad {\rm ,}
\end{equation}
\noindent with $h_{0} ,h_{1}$ constants. Normally $\vert h_{1}
\vert \ll a_{1} $, reflecting the weak dependence of $h/b$ on
temperature. It then follows from Eqs. (\ref{eq16}), (\ref{eq22}) and
(\ref{eq23}) that near the Sm-$A$--Sm-$C$ phase transition the
temperature dependence of the primary tilt is of the form
\begin{equation}
\label{eq24}
t^{2} \approx t_{0}^{2} (T_{A - C} - T) \quad {\rm ,}
\end{equation}
\noindent where the phase transition temperature $T_{A - C}$ is
given by
\begin{equation}
\label{eq25}
 T_{A - C} = T^{0} + h_{0} / (a_{1} - h_{1} ) \quad {\rm ,}
\end{equation}
and the constant scale factor in Eq. (\ref{eq24}) is $t_{0}^{2} =
a_{1} - h_{1}$.

\section{Sign inversion of the polarity order parameter}

Consider next the coefficient $R$ of Eq. (\ref{eq21}). If the
temperature dependence of all the coefficients entering the
expression for $R$ is neglected then, according to Eq.
(\ref{eq18}), the ratio $P_I / t$ would be constant with
temperature and it would follow from Eq. (\ref{eq24}) that $P_{I}
\sim \sqrt {T_{A - C}-T}$. However, it is apparent from Eq.
(\ref{eq21}) that this is not necessarily the case. Although each
of the coefficients is taken individually to vary slowly with
temperature, their combination could exhibit a rapid variation.
Specifically, the combination ${c}'{c}'' / {a}'$ represents the
couplings of the primary tilt and of the polarity to the secondary
tilt, scaled by the coefficient of the entropic contributions of
the latter. If this quantity is nearly equal to the coupling $c$
of the polarity to the primary tilt then the numerator on the
right hand side of Eq. (\ref{eq21}) would be very sensitive to the
temperature dependence of these two, mutually cancelling, terms.
The effect could be further magnified by the reduction of the
magnitude of the denominator if ${c'}^{2}$, associated with the
coupling of polarity to the secondary tilt, is not small compared
to the product ${a}'d$ associated with the entropic contribution
of these two parameters. Stated more briefly, $R$ is sensitive to
the relative strength of the coupling of the polarity to the
primary and secondary tilt. Two extreme situations can be
considered, corresponding to the complete decoupling of the
polarity from one of the tilt parameters. Thus if $P_{I}$ is
completely decoupled from ${t}'$, i.e. if ${c}' = 0$, then from
Eq. (\ref{eq21}) $R = c / d$. In the other extreme, polarity is
exclusively coupled to the secondary tilt, i.e. $c = 0$ and then
$R = - {c}'{c}'' / ({a}'d - {c'}^{2})$. In either of the decoupled
cases, the temperature dependence of $R$ does not involve mutually
cancelling terms and is weaker than the in fully coupled case.

To relate these considerations to the possible temperature
dependence of the $P_{I}$ we note that, quite generally, the
temperature dependence of $R$ may be approximated near the
transition temperature by
\begin{equation}
\label{eq26}
 R \approx R_{0} + R_{1} (T - T_{A - C} )\quad {\rm ,}
\end{equation}
\noindent where $R_{0} ,R_{1}$ are constants. Now, for the cases
where, as described above, $R_{1}$ is not negligible relative
to $R_{0}$, it is useful to define a characteristic
polarity-inversion temperature by
\begin{equation}
\label{eq27}
T_{inv}^{P} \equiv T_{A - C} - R_{0} / R_{1} \quad {\rm .}
\end{equation}
Near the transition temperature, $R$ can be expressed in terms of
$T_{inv}^{P}$ as
\begin{equation}
\label{eq28}
R \approx R_{1} (T - T_{inv}^{P} ) \quad {\rm .}
\end{equation}
It then follows from Eqs. (\ref{eq18}),(\ref{eq24}) and (\ref{eq28})
that the temperature dependence of $P_I$ is given by
\begin{equation}
\label{eq29}
P_{I} = t_{0} R_{1} (T - T_{inv}^{P} )\sqrt {T_{A - C} - T} \quad
{\rm .}
\end{equation}

\begin{figure}
  \centering
  \includegraphics*[width=8.6cm]{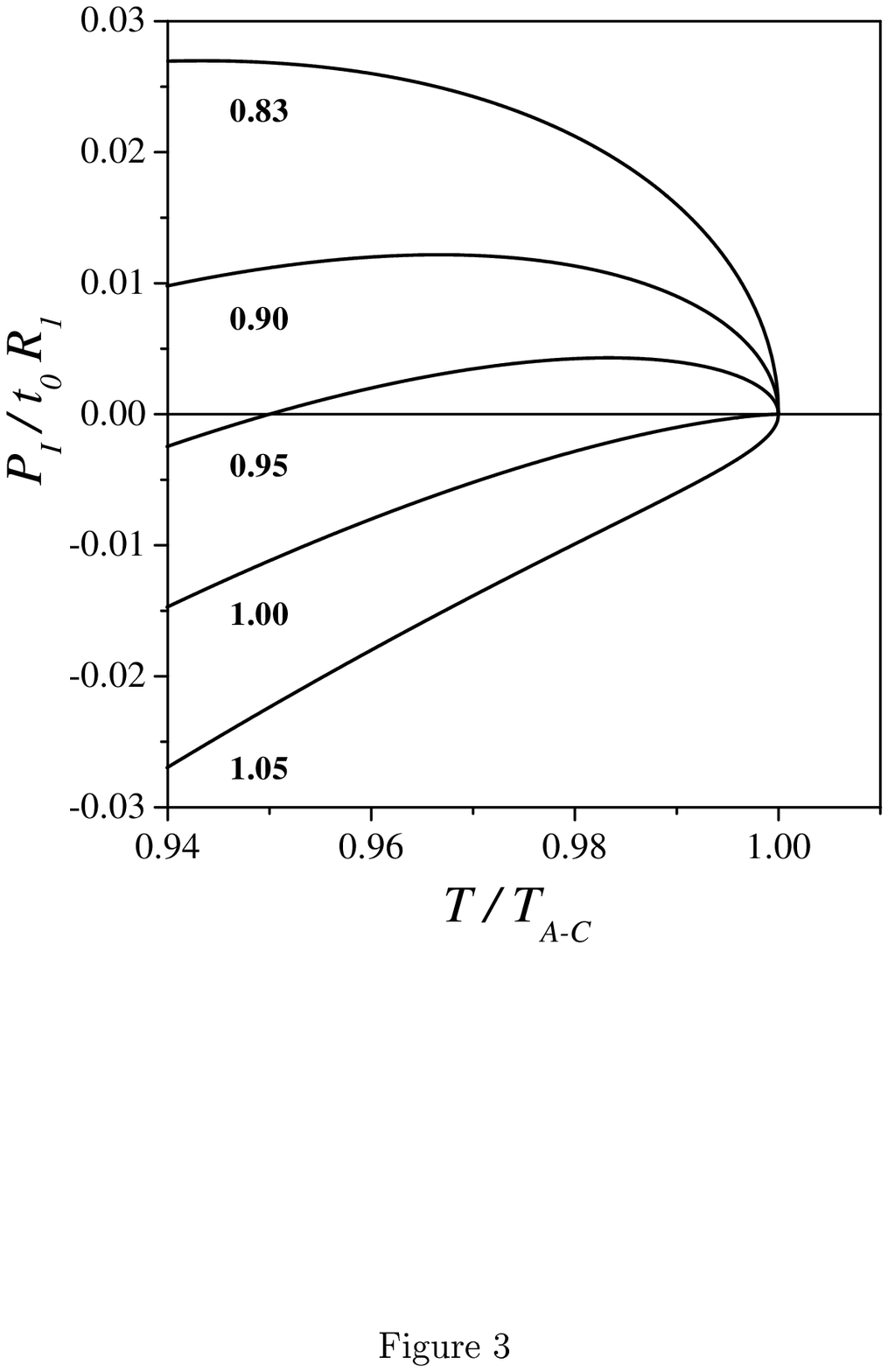}
\caption { Plots of the temperature dependence of the indigenous
polarity order parameter $P_{I}$ as calculated from Eq.
(\ref{eq29}) for different values of the ratio $T^{P}_{inv} / T_{A
- C}$ (printed on the left--hand end of each curve).} \label{fig3}
\end{figure}

Depending now on the value of $T_{inv}^{P} $ relative to $T_{A -
C}$, the polarity order parameter could exhibit either a
monotonous increase with decreasing temperature (if $T_{inv}^{P} >
T_{A - C}$) or a sign inverting variation (if $T_{inv}^{P} < T_{A
- C}$). In the latter case, the magnitude $P_{I}$ starts out from
zero at $T_{A - C}$ and increases continuously, on lowering the
temperature, to a local maximum at $T = (2T_{A-C}-T_{inv}^{P}
)/3$, then decreases until it vanishes at $T_{inv}^{P}$ and then
grows monotonously with inverted sign. Naturally, for the sign
inversion to be actually observed the inversion temperature
$T_{inv}^{P}$ should be lower than the transition temperature
$T_{A - C}$ but still within the temperature range of the Sm-$C$
phase. If $T_{inv}^{P}$ is too low, falling well outside the
range of the phase, then only the first part of the sign inverting
pattern, i.e. the continuous increase towards a local maximum, is
realised within the Sm-$C$ temperature range and this behaviour
appears qualitatively the same as the purely non--inverting
behaviour ($T_{inv}^{P} > T_{A - C}$). Altogether, according to
the result obtained in Eq. (\ref{eq29}) the different types of
temperature dependence can be classified according to the value of
a single characteristic parameter, the ratio $T_{inv}^{P} / T_{A -
C}$. The various possible cases according to this classification
are shown in Fig.~\ref{fig3}.

According to the relation in Eq. (\ref{eq14}), comparison of the
predicted temperature dependence of the indigenous polarity order
parameter $P_{I}$ with experiment is possible in the Sm-$C^{\ast}$
phase through measurements of the spontaneous polarisation
$P_{S}$. To use this relation it is necessary to specify the
``transverse dipole'' component $\mu _{\bot}^{*}$, and in
particular its dependence on temperature. This in turn depends on
the choice of the pseudovector $\mathbf{a}$ with respect to which
the indigenous polarity is defined according to Eq. (\ref{eq13}).
For rigid molecules, the orientation of $\bm \mu$ relative to
$\mathbf{a}$ will be fixed and therefore $\mu_{\bot}^{*}$ will be
strictly temperature independent. For flexible molecules, the
temperature dependence of $\mu _{\bot}^{*}$ will differ for
different choices of $\mathbf{a}$. In that case $\mu_{\bot} ^{*}$
will be temperature independent only if $\mathbf{a}$ is taken to
be fixed relative to the molecular segments to which the dipole
moment $\bm \mu$ is attached. In any case, assuming that
$\mathbf{a}$ is chosen in such a way that $\mu _{ \bot}^{*}$ does
not change appreciably with temperature over the range of the
Sm-$C^{\ast}$ phase, it follows from Eq. (\ref{eq14}) and
(\ref{eq29}) that the temperature dependence of the spontaneous
polarisation is of the form
\begin{equation}
\label{eq30}
P_{S} = P_{S}^{0} (T - T_{inv}^{P} )\sqrt {T_{A - C} - T} \quad {\rm
,}
\end{equation}
\noindent where $P_{S}^{0}$ is a temperature independent scale
factor. Figure 4 shows the comparison of the theoretical
temperature dependence with experimental measurements for
compounds exhibiting the temperature inverting
behaviour~\cite{bib9} as well as for compounds with the usual
monotonous variation of the spontaneous polarisation~\cite{bib30}.
The agreement is in all cases quite good and shows that the
classification of the different types of behaviour of the
compounds according to the single parameter $T_{inv}^{P}/T_{A- C}$
is quantitatively successful as well. Interestingly, the inversion
temperature $T^{P}_{inv}$ is found to be below the transition
temperature $T_{A-C}$, both for the sign inverting compounds and
for the monotonous one. Accordingly, the sign inversion in the
latter is precluded by the termination of the Sm-$C^*$ phase at a
temperature above $T^{P}_{inv}$. Moreover, the apparent monotonous
variation in this case is due to $T^{P}_{inv}$ being relatively
far below the Sm-$C^*$ temperature range (see top curve in
Fig~\ref{fig3}).

\begin{figure}\label{fig4}
  \centering
  \includegraphics*[width=8.6cm]{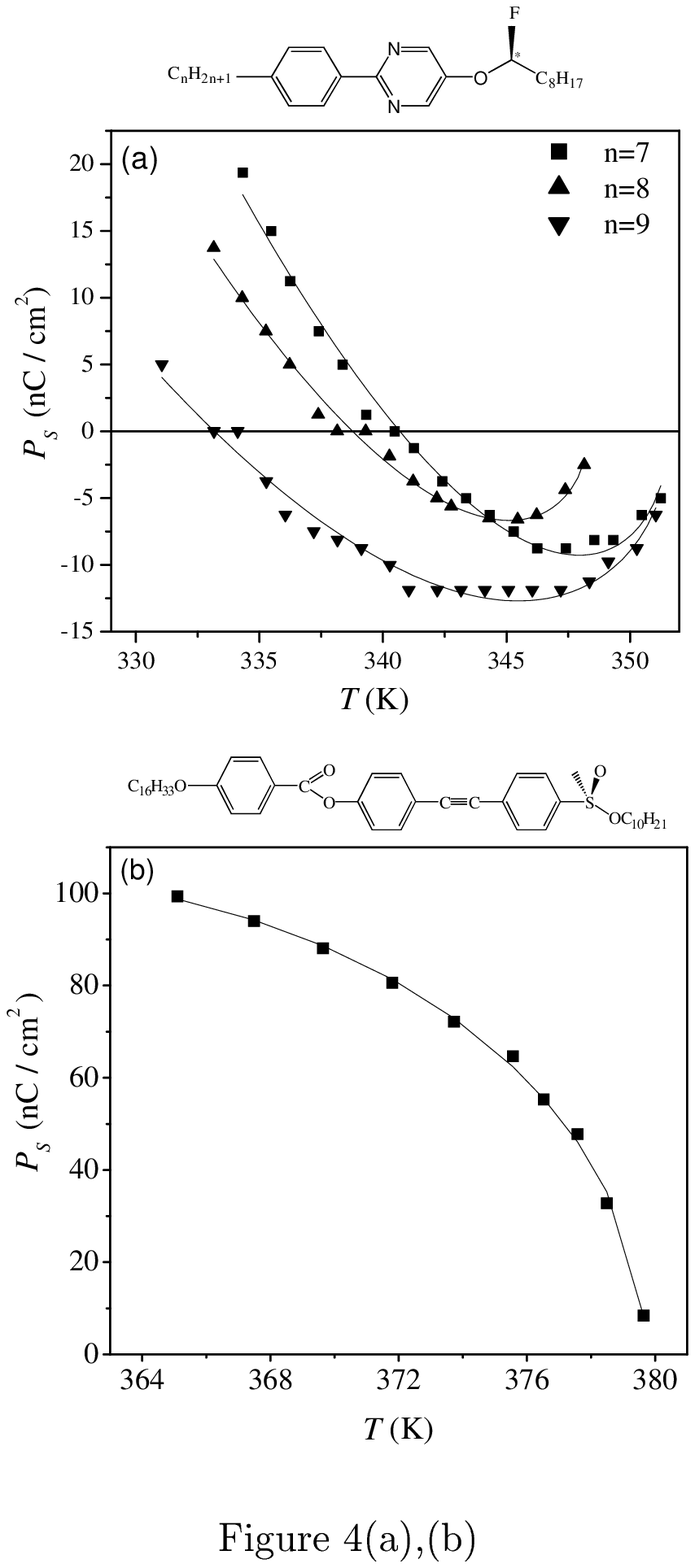} \\
\caption{Comparison of the theoretical temperature dependence of
the spontaneous polarisation $P_S$ with measurement. The chemical
structures of the compounds are drawn at the top of the respective
graphs. The continuous lines are theoretical fits according to
Eq.~(\ref{eq30}). (a)  Compounds exhibiting sign inversion (
experimental data from Ref. [9])  (b) Compound with the usual
monotonic variation of the spontaneous polarisation ( experimental
data from Ref. [30]). The characteristic temperature $T^{P}_{inv}$
for this compound is below the transition temperature $T_{A-C}$ by
$71K$. }
\end{figure}

It is apparent from the structure of the sign inverting compounds
in Fig.~\ref{fig4}(a) that the dipole moment (and chiral centre)
is situated right at the linkage of the mesogenic core to the
chiral end--chain and its ordering is therefore affected equally
strongly by the core and the tail. This is in accord with the
proposed mechanism of competing couplings of the polarity to the
primary and secondary tilt. Finally, it is worth noting here that
the different coupling of the polarity to the primary and the
secondary tilt makes it possible to differentiate the $P_{S}$
response of compounds that differ with respect to the position of
the transverse dipole moment within the molecular frame but are
otherwise similar in structure and therefore their tilts are
similar.

\section{Sign inversion of the secondary tilt.}

Analogous considerations apply to the possibility of sign
inversion in the temperature dependence of the secondary tilt. By
analogy to Eq. (\ref{eq26}), the temperature dependence of the
parameter $r$ of equations (\ref{eq17}), (\ref{eq20}) near the
transition can be expressed as
\begin{equation}
\label{eq31}
r \approx r_{0} + r_{1} (T - T_{A - C} ) \quad {\rm .}
\end{equation}
If the constant $r_{1}$ is negligible compared to $r_{0}$ the
ratio of tilts $t' / t$ in Eq. (\ref{eq17}) is temperature
independent. Otherwise, an inversion temperature $T_{inv}^{t'}$
for the secondary tilt can be defined in terms of the constants
$r_{0}$ and $r_{1}$
\begin{equation}
\label{eq32}
T_{inv}^{{t}'} = T_{A - C} - r_{0} / r_{1} \quad {\rm .}
\end{equation}
Combining Eqs. (\ref{eq17}), (\ref{eq24}) and (\ref{eq32}), the
following expression is obtained for the temperature dependence of
the secondary tilt near the phase transition
\begin{equation}
\label{eq33}
{t}' = t_{0} r_{1} (T - T_{inv}^{{t}'} )\sqrt {T_{A - C} - T} \quad
{\rm .}
\end{equation}
Accordingly, if $T_{inv}^{t'}$ falls within the temperature
range of the Sm-$C$ phase the temperature dependence of the
secondary tilt will exhibit a continuous sign inversion at
$T_{inv}^{t'}$. On comparing Eqs. (\ref{eq20}), (\ref{eq31})
and (\ref{eq32}) to the analogous set of Eqs. (\ref{eq21}),
(\ref{eq26}) and (\ref{eq27}) it becomes evident that the
inversion temperature of the secondary tilt, $T_{inv}^{t'}$, is
in general different from the inversion temperature of the
polarity $T_{inv}^{P}$. In particular, occurrence of one type of
inversion does not necessarily imply the occurrence of the other.

The vanishing and sign inversion of the secondary tilt with
temperature can be related the unwinding of the helix and
subsequent winding in opposite sense observed in some
Sm-$C^{\ast}$ compounds. This interpretation is based on the
assumption that the directions of the tilt vectors in adjacent
layers are correlated primarily through the direct interaction of
the flexible pendant chains on either side of the interface. If
then the secondary tilt $t'$ is identified with the effective tilt
order parameter of the end--chains, a sign inversion in $t'$ would
induce an inversion in the sense of the helical winding of the
primary tilt vector across the smectic layers.

\section{Reduction to a single tilt order parameter description and
comparison with the conventional theory}
The minimisation of the free energy in Eq. (\ref{eq15}) with
respect to $t'$ leads to the condition
\begin{equation}
\label{eq34}
{a}'{t}' = {c}'P_{I} - {c}''t \quad {\rm .}
\end{equation}
This condition can be used to eliminate the secondary tilt order
parameter from the original free energy expansion. The
${t}'$-minimised expression for the free energy, obtained from
Eqs. (\ref{eq15}) and (\ref{eq34}), has the form
\begin{equation}
\label{eq35}
\bar {g}_{A - C} = {\frac{{1}}{{2}}}\bar {a}t^{2} +
{\frac{{1}}{{4}}}bt^{4}
- \bar {c}P_{I} t + {\frac{{1}}{{2}}}\bar {d}P_{I}^{2} \quad {\rm ,}
\end{equation}
\noindent with
\begin{equation}
\label{eq36}
\begin{gathered}
  \bar a \equiv a - (c'')^2 /a' \hfill \\
  \bar c \equiv c - c'c''/a' \hfill \\
  \bar d \equiv d - (c')^2 /a' \hfill \\
\end{gathered}
 \quad {\rm .}
\end{equation}
Although the reduced Landau expansion in Eq. (\ref{eq35}) is
formally an expansion in $t$ and $P_{I}$, part of the information
associated with the eliminated secondary tilt ${t}'$ is implicitly
contained in the ``renormalized'' expansion coefficients through
their expressions in terms of the original coefficients as shown
in Eqs. (\ref{eq36}). In particular, as discussed in Sec. IV, the
renormalized coefficients $\bar{c}$ and $\bar{d}$ could become
sensitive to temperature variations around the Sm-$A$--Sm-$C$
transition in spite of the relative insensitivity of the
individual coefficients of the original expansion that combine to
produce them. However, when the reduced expansion is considered as
the starting point of the description, such sensitivity to
temperature can only be introduced \textit{add hoc}.

Mathematically, the form of the reduced expansion is identical to
the conventional Landau expansion, in its minimal form, used for
the free energy of a single layer of the Sm-$C^{\ast}$
phase~\cite{bib1,bib26,bib27,bib28}, namely
\begin{equation}
\label{eq37}
g_{A - C}^{\ast}  = {\frac{{1}}{{2}}}at^{2} + {\frac{{1}}{{4}}}bt^{4}
-
CP_{S} t + {\frac{{1}}{{2\varepsilon _{0} \chi _{0}} }}P_{S}^{2}
\quad {\rm
.}
\end{equation}
To a large extent, however, the resemblance is only formal as the
underlying physics is different. Equation (\ref{eq37}) describes
chiral compounds; in the absence of chirality it reduces to Eq.
(\ref{eq8}). The coefficient $C$ is assumed to be a pseudoscalar
associated in some way with molecular chirality. The quadratic
term in $P_S$ is taken to represent the entropic contribution
associated with the ordering of the molecular dipoles.
Accordingly, in the case of achiral compounds no such contribution
is allowed by the conventional theory. By contrast, the expansion
of Eq. (\ref{eq35}) takes into account the indigenous polarity and
therefore admits such entropic contributions for both chiral and
achiral tilted smectic phases. Since the polarity is included
irrespective of molecular chirality, in the case of chiral
molecules the free energy in Eq. (\ref{eq35}) is modified only to
the extent dictated by the additional interactions associated with
molecular chirality and the electrostatic forces among the
transverse molecular dipoles. Normally the effects of such
interactions on the stability of tilted and polar ordering are
estimated to be rather marginal. Thus the free energy would
include a direct electrostatic contribution of the form
\begin{equation}
\label{eq38}
 - d_{el} P_{S}^{2} = - d_{el} (\mu _{\bot}^{*})^{2}P_{I}^{2}
\quad {\rm ,}
\end{equation}
\noindent with the coefficient $d_{el} > 0$ and with the magnitude
of this electrostatic term much smaller than that of the entropic
term, i.e.
\begin{equation}
\label{eq39}
\lambda \equiv d_{el} (\mu _{ \bot} ^{ *}  )^{{\rm {\bf 2}}} / \bar
{d} \ll
1 \quad {\rm .}
\end{equation}
The differences in the physics underlying the Landau expansions in
Eqs. (\ref{eq35}) and (\ref{eq37}) have direct implications on the
thermodynamics of the Sm-$A$--Sm-$C$ transition. For example,
ignoring all inter-layer (helical structure etc) contributions to
the free energy, Eq. (\ref{eq37}) gives the following expression
for the difference between the transition temperatures of the
chiral (pure enantiomer) and achiral (racemic) phases~\cite{bib1}
\begin{equation}
\label{eq40}
 T_{A - C^\ast}  - T_{A - C} = (\varepsilon _{0} \chi
_{0} / a_{0} )C^{2} \quad {\rm ,}
\end{equation}
where the temperature dependance of the parameter $a$, near the
phase transition, is taken to be $a \approx a_0(T-T^0)$. The
result obtained for this difference from the reduced expansion of
Eq. (\ref{eq35}) is
\begin{equation}
\label{eq41}
 T_{A-C^\ast}- T_{A - C} = (\bar {c}^{2} / \bar
{d})(\lambda / (1 - \lambda )) \quad {\rm ,}
\end{equation}
\noindent and is essentially proportional to the rather small
relative contribution $\lambda$ of the electrostatic interactions
associated with the molecular dipole moment components that
survive as a result of the chiral asymmetry of the molecules. The
smallness of the predicted temperature shift is in agreement with
the rather small values generally obtained from measurements on
enantiomeric mixtures~\cite{bib31,bib32}. For a direct
quantitative comparison, however, it would be necessary to take
into account the contributions associated with the helical winding
of the director across the smectic layers of the chiral phase.

It has been suggested~\cite{bib16} that an additional, higher
order, ``piezoelectric'' term ${C}'P_{S} t^{3}$ should be included
in Eq. (\ref{eq37}) in order to account for the sign inverting
temperature dependence of $P_{S}$ in the context of the
conventional Landau expansion. The resulting free energy expansion
leads to the following dependence of the spontaneous polarisation
on the tilt
\begin{equation}
\label{eq42}
{\frac{{P_{S}} }{{t}}} = \varepsilon _{o} \chi _{o} (C - {C}'t^{2})
\quad {\rm .}
\end{equation}
Accordingly, $P_{S}$ would undergo a sign inversion at a
temperature where $t^{2}$ would become equal to $C /{C}'$. For
sign inversion to occur it is therefore required that $C$ and ${C}'$
be of the same sign (in the convention used here) and that ${C}'
\gg C$, since the expansion is valid for small $t^{2}$. Thus the
conventional description implies that the sign inverting compounds
belong to a class where, for some reason, the lower order
piezoelectric coefficient $C$ is much weaker than the higher order
one, ${C}'$, i.e. to a class of compounds that are in marked
contrast with the normally assumed ascending relative significance
of higher order expansion terms near the phase transition.

Under these conditions for $C$, and ${C}'$, the temperature
dependence for the spontaneous polarisation in the case of a
second order Sm-$A$--Sm-$C^{\ast}$ phase transition is of he form
\begin{equation}
\label{eq43}
P_{S} \sim (T - T_{inv}^{\ast}  )\sqrt {T_{A - C} - T} \quad {\rm ,}
\end{equation}
\noindent with the inversion temperature parameter given by
\begin{equation}
\label{eq44} T_{inv}^{\ast}  = T_{A - C} - C / {C}'a_{0} (b +
4C{C}'\varepsilon _{0} \chi_{0} ) \quad {\rm .}
\end{equation}
It is apparent, on comparing Eq. (\ref{eq43}) with Eq.
(\ref{eq30}), that the conventional Landau expansion with higher
order piezoelectric contributions and the extended expansion in
Eq. (\ref{eq15}), using the indigenous polarity and the secondary
tilt, lead to functionally identical forms for the temperature
dependence of $P_{S}$. Each form is parameterised by the
transition temperature $T_{A-C}$ and an inversion temperature.
However, the underlying physical picture is different and the
inversion temperatures are related to physically different
expansion coefficients: $T_{inv}^{P}$ of Eq. (\ref{eq30}) is
related to the coupling of the indigenous polarity to the primary
and the secondary tilt and applies to both chiral and achiral
molecules, whereas $T_{inv}^{\ast}$ applies only to chiral
molecules and is related to the piezoelectric coefficients $C$ and
${C}'$.

Finally, on further minimising the free energy in Eq. (\ref{eq35})
with respect to $P_{I}$, the indigenous polarity order parameter
can be eliminated from the expression of the minimised free
energy, yielding an expansion in only the primary tilt order
parameter $t$. This expansion has the same form with the
conventional expansion in Eq.~(\ref{eq8}). However, the
renormalized coefficients in the ($t',P_I$)--minimised expansion
are functions of the coefficients of the initial expansion of
Eq.~(\ref{eq15}) rather than ``starting" coefficients as in
Eq.~(\ref{eq8}).

\section{Discussion and Conclusions}
We have shown that the conventional phenomenological description
of the Sm-$A$--Sm-$C$ phase transition in terms of a single tilt
order parameter is applicable only to molecules that have an axis
of higher than two--fold rotational symmetry. Such molecular
symmetry requirements, however, are not met by any of the real
molecules forming Sm-$C$ phases. We have also shown that the
symmetries and the conformational structure of the real molecules
give rise to several, mutually independent, tilt order parameters
and also to polar arrangement of the molecules. The latter is
described by pseudovector ``indigenous polarity'' order parameters
and is shown to be compatible with the symmetries of the achiral
Sm-$C$ phase as well as of the chiral Sm-$C*$.

A phenomenological Landau expansion in terms of two tilt order
parameters (primary and secondary) and of an indigenous polarity
order parameter $P_I$ is shown to describe consistently the
Sm-$A$--Sm-$C$ transition and the appearance of spontaneous
electric polarisation $P_{S}$ in the chiral Sm-$C^{\ast}$. The
relation of $P_{S}$ to $P_{I}$ is established by means of a well
defined molecular quantity $\mu_{\bot}^\ast$ measuring the
electrostatic chirality of the molecule. The derived temperature
dependence of the spontaneous polarisation involves a single
characteristic reduced temperature $T_{inv}^{P} / T_{A-C} $ whose
value differentiates between the compounds showing the usual
monotonic variation of the spontaneous polarisation with
temperature and those exhibiting a sign inverting variation.
Temperature dependence measurements on both types of compounds are
accounted for very accurately.

On the phenomenological level, the sign inversion of $P_{S}$ is
obtained as a result of competition between the coupling of the
indigenous polarity to the primary tilt order parameter and to the
secondary one. On the molecular level, the implications of this
competition are compatible with the picture of competing molecular
conformations of opposite contributions to the spontaneous
polarisation. They do not exclude, however, the picture of
competing intermolecular interactions, particularly if the
conformational changes affect substantially the global structure
of the molecule, not just the part that contributes to the
spontaneous polarisation.

A similar sign inverting behaviour is found possible for the
secondary tilt order parameter and can be related to the inversion
of the helical pitch. In the underlying molecular picture the
secondary tilt is associated with the tail segments, which
essentially control the intra-layer correlations of the primary
tilt.

The new, extended, description differs from the conventional one
mainly in that it recognises that (i) polar ordering is present in
the tilted smectic phase and is not a result of chirality and (ii)
the tilted ordering is not always adequately described in terms of
a single order parameter. On eliminating, by minimisation of the
free energy, the secondary tilt and the indigenous polarity order
parameter, the extended Landau expansion reduces to the
conventional form of the expansion for the Sm-$A$ to Sm-$C$ (or
Sm-$C^{\ast}$) transition but with different physical content for
the expansion coefficients.

\begin{acknowledgments}
This work was supported in part by the Greek General Secretariat
of Research and Technology and the European Social Fund under the
PENED'99 project 99ED52.\\
P.K.K. gratefully acknowledges financial support through a
research scholarship from the University of Patras under project
"Karatheodori Scientific Research Program", grant No.1931.\\
D.J.P. thanks Ed Samulski for many stimulating discussions on the
subject of this paper.
\end{acknowledgments}

\appendix
\section{}
\begin{figure}
  \centering
  \includegraphics*[width=4cm]{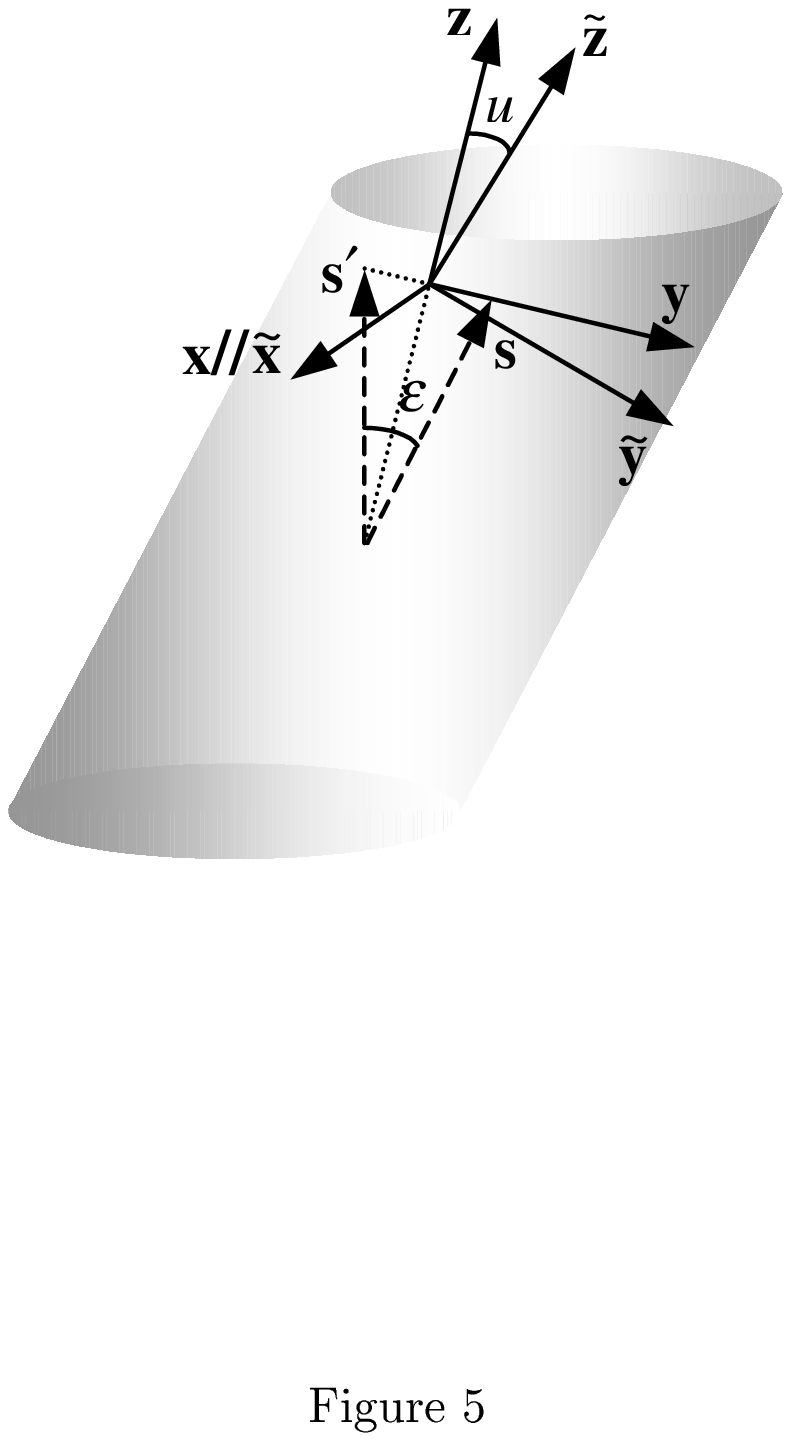}
\caption { Graphical representation of the molecular vectors, axis
frames and angles introduced in Equations (A1) and (A8) in
relation to the oblique cylinder geometry.} \label{figA1}
\end{figure}

The purpose of this appendix is to identify different sets of
independent order parameters describing tilt and polarity and to
determine the relations among such sets. This is done for rigid
molecules possessing a plane of symmetry and a twofold rotation
axis perpendicular to it. For concreteness, we use the molecular
geometry in Fig.~\ref{fig1}(c) i.e. molecules whose orientation can be
described completely in terms of two non- collinear unit vectors
$\mathbf{s}$ and $\mathbf{s'}$ lying on the plane of
symmetry of the molecule and forming a fixed angle $\varepsilon$
(see Fig.~\ref{figA1}). As explained in section II.2, the relevant
orientational
order parameters up to second rank for such molecules are $\eta
_{ab}^{(s)}$, $\eta _{ab}^{({s}')}$, $\eta _{ab}^{(s,{s}')}$ and
${\left\langle {a_{x}}  \right\rangle}$ and they are given in
Eqs. (\ref{eq9})-(\ref{eq12}).

We define an orthogonal frame of molecular axes $xyz$ such that $x$
coincides with the two-fold symmetry axis and the unit vectors
along the other two axes are related to $\mathbf{s}$ and
$\mathbf{s'}$ as follows,
\begin{eqnarray}
\begin{gathered}
  \mathbf{z} = (\mathbf{s} + \mathbf{s'})/2\cos (\varepsilon
/2) \hfill \\
  \mathbf{y} = (\mathbf{s} - \mathbf{s'})/2\sin (\varepsilon
/2) \hfill \\
\end{gathered} \quad {\rm .}
\end{eqnarray}
By analogy then with the order parameters in Eqs.
(\ref{eq9})-(\ref{eq11}) we can define the order parameters
associated with the unit vectors of the molecular frame
\begin{equation}
\begin{gathered}
  \eta _{ab}^{(z)}  = (3 < z_a z_b  >  - \delta _{ab} )/2
\hfill \\
  \eta _{ab}^{(y)}  = (3 < y_a y_b  >  - \delta _{ab} )/2
\hfill \\
  \eta _{ab}^{(y,z)}  =  3< z_a y_b  + y_a z_b  > /4
\hfill \\
\end{gathered}\quad {\rm .}
\end{equation}
The polar order parameter ${\left\langle {a_{x}} \right\rangle}$
is expressed in terms of the molecular axes as
\begin{equation}
\left\langle {a_X } \right\rangle  = \sin \varepsilon \left\langle
{(\mathbf{z} \times \mathbf{y})_X } \right\rangle  =  - \sin
\varepsilon \left\langle {x_X } \right\rangle \quad {\rm .}
\end{equation}
The second rank order parameters associated with the vectors
$\mathbf{s}$, $\mathbf{s'}$ can be obtained from the
respective order parameters of the molecular frames according to
the relations
\begin{equation}
\begin{gathered}
  \eta _{ab}^{(s)}  = \cos ^2 (\varepsilon /2)\eta _{ab}^{(z)} +
\sin ^2 (\varepsilon /2)\eta _{ab}^{(y)}  + \sin \varepsilon \eta
_{ab}^{(y,z)} \hfill \\
  \eta _{ab}^{(s')}  = \cos ^2 (\varepsilon /2)\eta _{ab}^{(z)} +
\sin ^2 (\varepsilon /2)\eta _{ab}^{(y)}  - \sin \varepsilon \eta
_{ab}^{(y,z)} \hfill \\
  \eta _{ab}^{(s,s')}  = \cos ^2 (\varepsilon /2)\eta _{ab}^{(z)} -
\sin ^2 (\varepsilon /2)\eta _{ab}^{(y)}  \hfill \\
\end{gathered}\quad {\rm .}
\end{equation}
Now, if we describe the orientation of the molecular frame $xyz$
relative to the macroscopic, phase fixed, frame $XYZ$ by
the three Euler angles~\cite{bib33} $\phi ,\vartheta ,\psi$, the
relevant tilt and polarity order parameters associated with the
molecular axis frame are given by
\begin{equation}
\begin{gathered}
  \eta _{YZ}^{(z)}  =  -  < \sin 2\vartheta \cos \phi  > /2 \hfill
\\
  \eta _{YZ}^{(y)}  =  < \sin 2\vartheta \cos \phi \cos ^2 \psi -
\sin \vartheta \sin \phi \sin 2\psi  > /2 \hfill \\
  \eta _{YZ}^{(yz)}  =  < \cos 2\vartheta \cos \phi \cos \psi  - \cos
\vartheta \sin \phi \sin \psi  > /2 \hfill \\
   < x_X  >  =  < \cos \vartheta \sin \phi \sin \psi  - \cos \phi
\cos \psi  > /2 \hfill \\
\end{gathered} \quad {\rm .}
\end{equation}
It is apparent from these equations that the four order parameters
are in general independent. In the special case where the $\psi$
rotations are completely unbiased (higher than two-fold rotational
symmetry about the molecular $z$-axis), only one independent tilt
order parameter survives since then Eqs. (A5) yield
\begin{equation}
\eta _{YZ}^{(y)} = - \eta _{YZ}^{(z)} / 2 \quad {\rm ,}
\end{equation}
\noindent and
\begin{equation}
\eta _{YZ}^{(yz)} = 0 = {\left\langle {x_{X}}  \right\rangle}
\quad {\rm .}
\end{equation}
In the general case, it is always possible to eliminate the
``mixed'' order parameter $\eta _{YZ}^{(yz)}$ by rotating the
molecular frame about the $x$-axis by an angle $u$ to the
molecular frame of the principal molecular axes
$\tilde{x},\tilde{y},\tilde {z}$ (see Fig.~\ref{figA1}). The angle
$u$ of rotation is given by the relation
\begin{equation}
\tan 2u = {{2\eta _{YZ}^{(yz)} } \mathord{\left/
 {\vphantom {{2\eta _{YZ}^{(yz)} } {(\eta _{YZ}^{(z)}  - \eta
_{YZ}^{(y)} )}}} \right.
 \kern-\nulldelimiterspace} {(\eta _{YZ}^{(z)}  - \eta _{YZ}^{(y)}
)}} \quad {\rm .}
\end{equation}
This rotation makes the mixed order parameter vanish and leaves
the polar order parameter invariant since the axes $x$ and $\tilde
{x}$ coincide. The order parameters expressed in the two frames
are related as follows,
\begin{equation}
\begin{gathered}
  \eta _{YZ}^{(z)}  = \eta _{YZ}^{(\tilde z)} \cos ^2 u \hfill \\
  \eta _{YZ}^{(y)}  = \eta _{YZ}^{(\tilde y)} \sin ^2 u \hfill \\
  \eta _{YZ}^{(yz)}  = (\eta _{YZ}^{(\tilde z)}  - \eta
_{YZ}^{(\tilde y)} )\sin 2u/2 \hfill \\
  \eta _{YZ}^{(\tilde y\tilde z)}  = 0 \hfill \\
  \left\langle {\tilde x_X } \right\rangle  = \left\langle {x_X }
\right\rangle  \hfill \\
\end{gathered}\quad {\rm .}
\end{equation}
Accordingly, it is possible to replace the description in terms of
three tilt order parameters (associated with the tensor components
$\eta _{YZ}^{(s)} ,\eta _{YZ}^{({s}')} ,\eta _{YZ}^{(s,{s}')}$) by
a description in terms of just two tilt parameters (associated
with $\eta _{YZ}^{(\tilde {z})} ,\eta_{YZ}^{(\tilde {y})}$) and a
molecular axis rotation angle $u$. The polar order parameter is
identical in both descriptions.

\end{multicols}
\end{document}